\documentclass[%
 twocolumn,
 superscriptaddress,
 nofootinbib,
 amsmath,amssymb,amsfonts
 aps,
prb,
 floatfix,
]{revtex4-2}

\usepackage{color}
\usepackage{enumerate}
\usepackage{graphicx}
\usepackage{changes}
\usepackage[breaklinks=true,colorlinks,citecolor=blue,linkcolor=blue,urlcolor=blue]{hyperref}

\begin{document}

\title{The role of correlated hopping in many-body physics of flat-band systems: Nagaoka ferromagnetism}

\author{Tom Westerhout}%
  \email{tom.westerhout@ru.nl}%
\author{Mikhail I. Katsnelson}%
 \email{mikhail.katsnelson@ru.nl}%
  \affiliation{Institute for Molecules and Materials, Radboud University, Heijendaalseweg 135, 6525 AJ Nijmegen, The Netherlands}

\date{\today}

\begin{abstract}
In narrow-band systems, correlated contribution to effective hopping becomes important, and one needs to carefully consider three types of hopping processes: between doubly and singly occupied sites, between singly occupied and empty sites, and between doubly occupied and empty or two singly occupied sites. All three hopping parameters cannot vanish simultaneously, and one should specify for which of these processes the band becomes flat. By means of exact diagonalization of finite systems we demonstrate that three hoppings play qualitatively different roles in many-body effects, in particular, in the formation of half-metallic ferromagnetic (Nagaoka) states.
\end{abstract}

\maketitle



In solids, the narrower the electronic band the more important the correlation effects are. A prototypical example of an electron system with a flat band is the quantum Hall regime for a two-dimensional electron gas in a magnetic field when the kinetic energy is suppressed by Landau quantization~\cite{quant_hall}. In this situation many-body effects become dominant leading, for example, to fractional quantum Hall effect, and all relevant terms in the Hamiltonian originate from the Coulomb interaction~\cite{murthy}. The conventional perturbative techniques, such as Feynman diagrammatic expansion, are then not applicable and more advanced and subtle theoretical approaches are required~\cite{hansson}. 

Flat bands can also be observed in solids even without a quantized magnetic field. The most well-known example is the magic-angle twisted bilayer graphene~\cite{morell2010flat,bistritzer2011moire,li2010observation} where the formation of a flat band is accompanied by a broad variety of strongly correlated effects and electronic phase transitions including unconventional superconductivity~\cite{cao2018correlated,cao2018unconventional,yankowitz2019tuning,sharpe2019emergent}. Another example is a single layer of InSe where an almost flat hole band arises without any twisting. The appearance of this band is accompanied, among other things, by an anomalously strong electron-phonon interaction and closeness to ferromagnetic and charge-density-wave instabilities~\cite{InSe1,InSe2}. Strong correlation effects are thus practically unavoidable in narrow-band systems.

Recently, it was emphasized~\cite{cea2021electrostatic} that in the flat-band regime, kinetic energy contribution from the Coulomb interaction, a so called \emph{correlated hopping}, becomes relevant.
This concept is very old, originating from the polar model of Shubin and Vonsovsky in 1934~\cite{schubin1934electron}. The polar model, including the correlated hopping effects, has since then been reconsidered with contemporary theoretical tools such as Hubbard X-operators in Refs.~\cite{vonsovsky1979some1,vonsovsky1979some2}. Independently, the concept was introduced in Ref.~\cite{foglio1979new}. A possible role of correlated hopping in the physics of high-temperature cuprate superconductivity was discussed by Hirsch~\cite{hirsch1989hole,hirsch1989bond}. In one dimension, the Hubbard-like model with correlated hopping can be solved exactly in some cases~\cite{arrachea1996anomalous,aligia2000phase}. The two-dimensional case has been considered numerically by means of exact diagonalization~\cite{gagliano1995single,arrachea2000pairing} and Quantum Monte Carlo~\cite{di2014quantum} methods. 

The discovery of real two-dimensional materials with flat (or almost flat) bands requires more careful and detailed studies of the correlation effects, one by one. In this Letter we focus on the appearance of a saturated, half-metallic ferromagnetic state typical for an almost half-filled narrow-band case. We call this phenomenon Nagaoka ferromagnetism due to the seminal work~\cite{nagaoka1966ferromagnetism}.
Nagaoka has proven rigorously that in the infinite-$U$ Hubbard model near half-filling, with just one extra or missing electron, the ground state is a saturated ferromagnet with maximum possible total spin. The result holds for various two- and three-dimensional crystal structures, but does not work in one dimension.
This initial formulation is, strictly speaking, insufficient since the gap between this ground state and excited states vanishes in thermodynamic limit. However, it was proven by Irkhin and Katsnelson via virial expansion for X-operator Green's functions~\cite{irkhin1985spin} that Nagaoka's result remains valid for finite but small concentrations of extra (or missing) electrons. Further increase in electron (or hole) concentration results in a transition to a non saturated ferromagnetic state~\cite{edwards,irkhin1990katsnelson,irkhin_zarubin}, but rigorous results about this transition are still missing.
The case of a one-dimensional system with next- and next-nearest-neighbor hoppings was considered in Ref.~\onlinecite{muller1995ferromagnetism}. Rigorous results for ferromagnetism in the Hubbard model on a special class of decorated lattices with degenerate ground states were obtained in Ref.~\onlinecite{mielke1993ferromagnetism}. Ferromagnetism in the Hubbard model was also studied using dynamical mean-field theory which is formally associated to the limit of infinite dimension~\cite{uhrig1996exact,held1996correlated,vollhardt1996non}. It is impossible to give a full list of references on the topic, but important works~\cite{tasaki1998nagaoka,maksymenko2012flat,mielke1999ferromagnetism,tamura2002flat,roth1969electron,jarrett1968evidence,plischke1974ferromagnetism,hanisch1997lattice} should be mentioned to illustrate different approaches to the problem.


We start our analysis with the standard Hubbard model with nearest-neighbor hopping:

\begin{equation}\label{eq:hubbard}
    H_\text{Hubbard} = t \sum_{\substack{\langle i,j\rangle, \;\sigma \\ i < j}} (c^\dagger_{i\sigma} c^{\vphantom{\dagger}}_{j\sigma} + \text{H.c.})
      + U \sum_{i} n_{i\uparrow}n_{i\downarrow} \,.
\end{equation}
Here, $c^\dagger_{i\sigma}$ and $c^{\vphantom{\dagger}}_{i\sigma}$ are fermionic creation and annihilation operators respectively, creating or annihilating an electron at site $i$ with spin $\sigma = \{ \uparrow, \downarrow \}$; $n_{i\sigma} = c^\dagger_{i\sigma} c^{\vphantom{\dagger}}_{i\sigma}$; $\langle i, j \rangle$ denote the nearest neighbors.

We focus on the almost half-filled case when the number of electrons is one fewer than the number of sites. For $U\to \infty$ the ground state of the system is ferromagnetic, at least, for some types of crystal lattices \cite{nagaoka1966ferromagnetism}. In the following, we will consider the effect of narrow or even flat bands on the stability of this ferromagnetic phase. 

\begin{figure*}[t]
    \centering
    \includegraphics[width=0.7\textwidth]{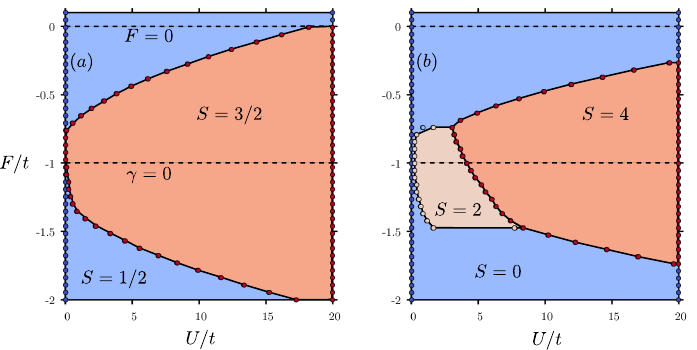}
    \caption{Phase diagrams of (a) $2\times 2$ plaquette with 3 electrons and (b) $3\times 3$ plaquette with 8 electrons as functions of $F$ and $U$. Points are numerical data and lines are a guide to eye. Different colors correspond to different values of the total spin $S$ defined in Eq.~\eqref{eq:spin}.}
    \label{fig:ferromagnetism}
\end{figure*}

To this end, instead of the Hamiltonian \eqref{eq:hubbard} with a single hopping parameter $t$, we consider a more complete polar model~\cite{schubin1934electron,vonsovsky1979some1,vonsovsky1979some2} which also contains hopping like terms originating from the electron-electron interaction. These terms contain three fermionic operators belonging to one site $i$ and one fermionic operator belonging to another site $j$. In the nearest-neighbor approximation they are described by one parameter,
\begin{equation}\label{eq:coulomb}
    F \!= \left\langle ii\left| v\right| ij\right\rangle \!= \!\!\int d{\bf r}d{\bf r}%
^{\prime }\psi _{i}^{\ast }({\bf r})\psi _{i}^{\ast }({\bf r}^{\prime
})v\left( {\bf r-r}^{\prime }\right) \psi _{i}({\bf r})\psi _{j}({\bf r}%
^{\prime }) \,,
\end{equation}
where $v\left( {\bf r-r}^{\prime }\right)$ is the potential of inter-electron interaction, $\psi _{i}({\bf r})$ is the state of the electron at site $i$ and the sites $i$, $j$ are nearest neighbors. 

After adding such terms to the Hamiltonian \eqref{eq:hubbard}, we obtain
\begin{multline*}
    H = 
        U \sum_{i} n_{i\uparrow}n_{i\downarrow}
        + t\!\!\sum_{\substack{\langle i,j\rangle, \;\sigma \\ i < j}}
        \!(c^\dagger_{i\sigma} c^{\vphantom{\dagger}}_{j\sigma} + \text{H.c.}) \\
        + F\!\!\sum_{\substack{\langle i,j\rangle, \;\sigma \\ i < j}}
        \!(n_{i\bar{\sigma}} + n_{j\bar{\sigma}})(c^\dagger_{i\sigma} c^{\vphantom{\dagger}}_{j\sigma} + \text{H.c.})\,.
\end{multline*}
Here, $\bar\sigma$ denotes the inverse of $\sigma$ such that if $\sigma=\,\uparrow$ then $\bar\sigma=\,\downarrow$, and if $\sigma=\,\downarrow$ then $\bar\sigma=\,\uparrow$. By considering different occupations of the sites connected by the hopping process, we can rewrite the Hamiltonian as
\begin{multline}\label{eq:extended}
    H \!= 
        U \sum_{i} n_{i\uparrow}n_{i\downarrow} +
        \!\!\sum_{\substack{\langle i,j\rangle, \;\sigma \\ i < j}}
        \!\bigg[ \beta'' (1 - n_{i\bar\sigma}) (1 - n_{j\bar\sigma}) +
              \beta' n_{i\bar\sigma} n_{j\bar\sigma} \\
            + \gamma \big(n_{i\bar\sigma} (1 - n_{j\bar\sigma}) + (1 - n_{i\bar\sigma}) n_{j\bar\sigma}\big) \bigg]
            (c^\dagger_{i\sigma} c^{\vphantom{\dagger}}_{j\sigma} + \text{H.c.}) \,,
\end{multline}
where
\begin{equation*}
\left\{ \begin{aligned}
    \beta'' &= t \\
    \beta' &= t + 2 F \\
    \gamma &= t + F
\end{aligned} \right.\,.
\end{equation*}
In the above, we have used the Wannier basis which is orthonormal by construction. The corresponding expressions in a more general case of non-orthogonal basis functions (for example, atomic orbitals) are derived in Ref.~\cite{irkhin}.

Generally speaking, the parameter $F$ is by no means small. Recently, it was estimated from the first-principle constrained random phase approximation (cRPA) method for benzene molecule~\cite{van2021random}. The results are: $F=0.53\;$eV, $t=-2.75\;$eV. For comparison, Hubbard $U$ for this system is $U=9.38\;$eV, nearest-neighbor Coulomb interaction is $V_\text{NN}=6.12\;$eV, and nearest-neighbor direct exchange interaction is $J=0.18\;$eV. Note that the parameters for benzene molecule are similar to those of graphene~\cite{schuler2013optimal}. 

The Hamiltonian (\ref{eq:extended}) is part of the complete polar model~\cite{vonsovsky1979some1} which originally also contained the intersite Coulomb interaction. Here $\beta''$, $\beta'$, and $\gamma$ are parameters describing hopping between an empty and a singly occupied site, between a doubly and singly occupied site, and between two singly occupied sites (or between doubly occupied and an empty site), respectively. To avoid confusion, note that our definition of the parameter $\beta''$ differs by a minus sign from the original one in Ref.~\cite{vonsovsky1979some1}. Importantly, at $F \neq 0$ the three hopping parameters cannot all be zero, but in the following we analyze the effect of zeroing one of them at a time. We will see that these three hopping parameters play different roles in the formation of magnetic ground state of the system under consideration. First, we present the numerical results and then give their qualitative explanation. 

We focus on the square lattice and perform exact diagonalization for small clusters to study how the total spin in the ground state depends on $t$, $F$, and $U$. The QuSpin package \cite{weinberg2019quspin} is used for exact diagonalization. The total spin is defined by
\begin{equation}\label{eq:spin}
    S(S + 1) = \sum_{ij} \mathbf{S}_i \cdot \mathbf{S}_j = \sum_{ij} (S^x_i S^x_j + S^y_i S^y_j + S^z_i S^z_j)\,,
\end{equation}
where
\begin{equation*}
    S^x_i = \frac{1}{2} \sum_{\sigma\sigma'} c^\dagger_{i\sigma'} \sigma^x_{\sigma'\sigma} c^{\vphantom{\dagger}}_{i\sigma}
\end{equation*}
and $S^y$ and $S^z$ are defined similarly. $\sigma^x$, $\sigma^y$, $\sigma^z$ are Pauli matrices.


Let us first consider a $2\times 2$ cluster. Even though the system is very small, it still reproduces an experimentally realizable situation of a quadruple quantum dot where Nagaoka ferromagnetic ground state was recently observed~\cite{dehollain2020nagaoka}. In Fig.~\ref{fig:ferromagnetism}(a) we show the phase diagram of the system. Since for three electrons there are only two possible values of the total spin, $1/2$ and $3/2$, we find two phases. The horizontal line at $F=0$ corresponds to the standard Hubbard model, and, as expected, the system is antiferromagnetic at $U=0$ and becomes ferromagnetic for $U/t \gtrsim 18.4$.

The main difference between the Hubbard model with $U=0$ and $U\to\infty$, is the ``Gutzwiller projector'' forbidding the appearance of doubly occupied sites~\cite{vollhardt}. Putting $\gamma=0$ has the same consequences as $U\to\infty$ since it makes impossible the creation (as well as annihilation) of doubly occupied and empty sites from a pair of singly occupied sites. This means that with $\gamma=0$ the system behaves as an infinitely strongly correlated one even for small $U$, and one could expect that Nagaoka ferromagnetism arises even if the Coulomb interaction is not large. We indeed observe this behavior in Fig.~\ref{fig:ferromagnetism}(a) where for $\gamma=0$ (in other words, at $F=-1$) ferromagnetism is favored even at very small $U$. At the same time, setting $F$ close enough to $0$, that is considering small $\beta''$, makes ferromagnetism impossible.

The simple example of a $2\times 2$ cluster already clearly shows qualitatively different roles of the hoppings parameters in Eq.~\eqref{eq:extended}: whereas small $\gamma$ favors ferromagnetism, small $\beta''$ destroys it. In Fig.~\ref{fig:ferromagnetism}(b) we show the phase diagram for a $3\times 3$ cluster with eight electrons. Here, the situation is very similar: for $\gamma=0$ and $U/t \gtrsim 4.2$, Nagaoka ferromagnetic ground state is realized, and for $\beta''=0$ the system is always antiferromagnetic. One additional feature which we observe for a larger system is the stabilization of a non-saturated ferromagnetic phase for small $U$ and $\gamma$. In the Supplemental Material~\cite{supplement} we also provide data for clusters of $5$, $6$, $8$, $10$, $12$, and $13$ sites---they all demonstrate qualitatively similar behavior.


Let us now discuss qualitatively the physical origin of Nagaoka ferromagnetism. First, note that $\beta'$ and $\beta''$ are dual with respect to particle-hole inversion. Hence in the following, without loss of generality, we discuss the role of $\beta''$ since in the numerical simulations we had one extra hole.

Nagaoka ferromagnetism can be understood in terms of effective narrowing of the electronic band when an electron moves in a non-ferromagnetic environment---the mechanism known as double exchange~\cite{anderson1963theory}. The band edges remain the same for any spin configuration, but the density of states drops exponentially in case of random or antiferromagnetic spin configurations, forming so called Lifshitz tails~\cite{brinkman1970single,auslender2005electron,auslender2006electron}. Thus, for any small but finite charge carrier occupation, the lowest average band energy is obtained for a ferromagnet. An explicit spin Hamiltonian describing this tendency can be derived in the static approximation for the narrow-band Hubbard model on Bethe lattice~\cite{auslender1982effective}; it confirms the previously known non-Heisenberg form of the corresponding exchange interaction which is proportional to $J\cos(\theta_{ij}/2)$, where $\theta_{ij}$ is the angle between magnetic moments at sites $i$ and $j$~\cite{anderson1963theory}.

Importantly, the coupling constant $J$ is actually $\beta''$ since this band narrowing effect is related to the hopping between singly occupied and empty sites. This explains the tendency to ferromagnetism at finite $\beta''$ and vanishing $\gamma$; the latter makes the system strongly correlated and the effective $U$ infinitely large, forbidding the appearance of doubly occupied sites. 

The role of finite $\gamma$ is the exact opposite. Creation and annihilation of doubly occupied and empty site pairs decrease the energy of the antiferromagnetic state resulting in indirect (``kinetic'') antiferromagnetic exchange~\cite{anderson1963theory}. Its value in the large-$U$ Hubbard model is proportional to $-t^2/U$ and in the full polar model it becomes $-\gamma^2/U$. Therefore, finite $\gamma$ favors antiferromagnetism. 

One charge carrier (either an electron or a hole) put in an antiferromagnetic environment forms a magnetic polaron (also known as fluctuon or ferron) creating around itself a local ferromagnetic environment~\cite{krivoglaz1973fluctuating,nagaev1983physics,nagaev2001colossal,visscher1974phase,mott1973metal,auslender1981magnetic,auslender2005electron,auslender2006electron,soriano2020magnetic}. In the simplest model of an infinite potential well in two dimensions~\cite{soriano2020magnetic}, the energy of formation of such a ferromagnetic droplet can be estimated as
\begin{equation*}
    \mathcal{E}(R) = -\Delta + \frac{\hbar^2 z_0^2}{2m^*R^2} + J\frac{\pi R^2}{S_0} \,,
\end{equation*}
where $\Delta$ is the energy difference between the cases of antiferromagnetic and ferromagnetic environments,
$R$ is the radius of the well, $z_0 = 2.40483$ is the first zero of the Bessel function, $J_0(z)$, $m^*$ is the effective mass, $J$ is the antiferromagnetic indirect exchange, and $S_0$ is the area per magnetic atom. After optimization with respect to $R$, we estimate the magnetic polaron energy as
\begin{equation*}
    \mathcal{E} = -\Delta + C\sqrt{WJ},
\end{equation*}
where $W$ is the total bandwidth and $C$ is some numerical factor (its value can be found in Ref.~\cite{soriano2020magnetic}). To be stable, the magnetic polaron should have negative energy. Assuming that both $\Delta$ and $W$ are proportional to $\beta''$ and that $J \propto \gamma^2/U$, one obtains the stability criterion
\begin{equation}\label{eq:eqn33}
    \gamma^2 < D|\beta''|U,
\end{equation}
where $D$ is another numerical factor. Eq.~\eqref{eq:eqn33} is too rough to be quantitatively correct, especially for small clusters, but qualitatively it does explain the opposite roles of $\beta''$ and $\gamma$---the first one stabilizing and the latter destabilizing Nagaoka ferromagnetism. 

To conclude, in this work we have analyzed the role of correlated hopping in narrow-band systems focusing on one particular many-body phenomenon---Nagaoka ferromagnetism. By performing exact diagonalization of small clusters we have shown that for one extra hole (electron) $\beta''$ ($\beta'$) favors the ferromagnetic ground state whereas $\gamma$ favors the antiferromagnetic phase. Smallness of $\gamma$ acts as the Gutzwiller projector making the system strongly correlated even at small values of the Hubbard $U$. We also provide qualitative explanation of our numerical results. 

The code to carry out the analysis is publicly available at \url{http://github.com/twesterhout/correlated-hoppings}.




This work is supported by European Research Council via Synergy Grant No. 854843 - FASTCORR. Numerical simulations in this work were carried out on the Dutch national \mbox{e-infrastructure} with the support of SURF Cooperative.


\bibliography{references.bib}



\end{document}


\title{Supplemental Information to the article ``The role of correlated hoppings in many-body physics of flat-band systems: Nagaoka ferromagnetism''}
\author{Tom Westerhout}%
\author{Mikhail I. Katsnelson}%

\maketitle

\begin{figure*}[h]
    \centering
    \includegraphics[width=0.95\textwidth]{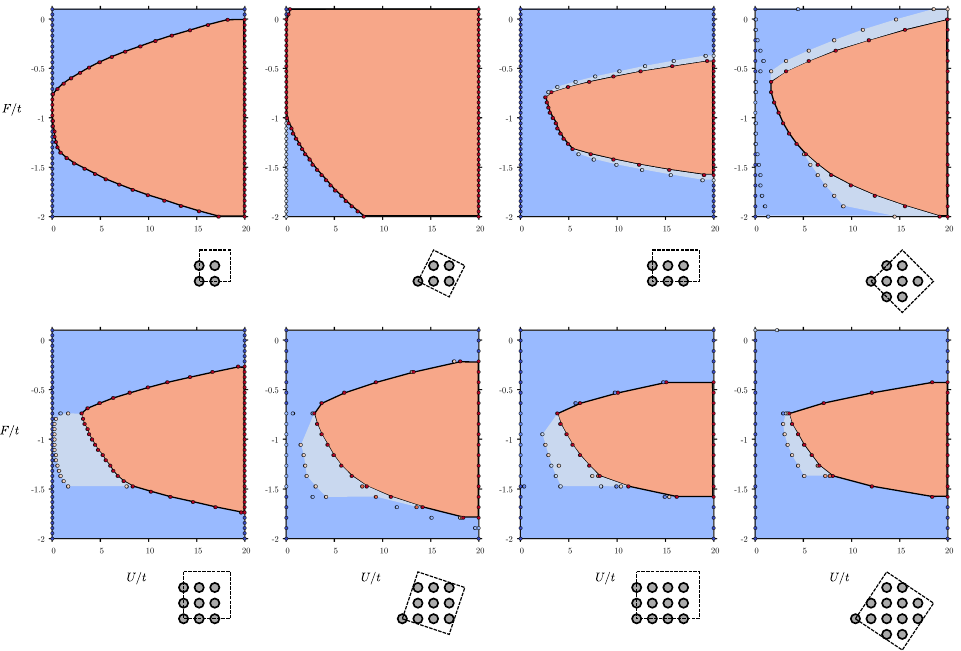}
    \caption{Phase diagrams as functions of $U/t$ and $F/t$ for clusters of various sizes. Points are numerical data and lines are a guide to eye. For each cluster we also show the shape of the unit cell in real space. Red regions indicate the saturated ferromagnetic phase, blue regions indicate the antiferromagnetic phase, and light blue -- phases with intermediate spin.}
    \label{fig:various-sizes}
\end{figure*}

In this document we provide numerical data for various cluster sizes. Specifically, in Figure~\ref{fig:various-sizes} we show phase diagrams similar to Figure~1 from the main text, but for clusters of 4, 5, 6, 8, 9, 10, 12, and 13 sites. In all cases we take the number of electrons to be one fewer than the number of sites.

